# Data Acquisition and Signal Processing for the Gamma Ray Energy Tracking Array (GRETA)

Thorsten Stezelberger*, *Member, IEEE*, John Joseph, Vamsi Vytla, Sergio Zimmermann

Lawrence Berkeley National Laboratory

*Abstract*— The Gamma Ray Energy Tracking Array (GRETA) is a 4-π detector system, currently under development, capable of determining energy, timing and tracking of multiple gamma-ray interactions inside germanium crystals as demonstrated in the Gamma Ray Energy Tracking In-Beam Array (GRETINA). Charge sensitive amplifiers instrument the crystals and their outputs are converted using analog to digital converters for real-time digital processing. In this paper, we will present the design of the detector system and data acquisition.  We will describe the real time components of the digital signal-processing path used to find the energy and timing of the gamma rays at low and high rates. We will describe the performance of the data acquisition system hardware and firmware and compare with the requirements.

## I. Introduction

THE GRETA detector, currently in the construction phase, builds on the experience of approximately seven years of GRETINA [1, 2] operations. The requirements of GRETA, when compared to GRETINA, include delivery of full sphere coverage (4π), increased gamma ray rate, improved gamma ray energy linearity and resolution and implementation of a modular design for future upgrades. The main GRETA systems are detector modules, data acquisition (DAQ) electronics, computing and the mechanical support structure. The DAQ electronics (Fig. 1) provides real time signal processing and data reduction.  It can be divided into four main hardware units: (a) 120 germanium crystals and charge sensitive amplifiers packaged into 30 Quad Detector Module. (i.e., four crystals per Quad), (b) Digitizer Modules (DM) which convert the analog detector signals into a digital data stream, (c) Signal Filter Boards (SFB) which accept triggers to reduce the data rate and extract the timing and energy information and (d) the computing systems which compute the gamma ray interaction points and tracks and build the events before saving the data to storage. All this happens in real-time, as the data flows through the system.

## II. Data Acquisition Electronics

The DMs are attached directly to detector modules to minimize noise and to preserve signal integrity (Fig. 2).  The DM digitizes the crystal core and 36 segment contacts using 40 100MSample/s Analog to Digital Converters (ADCs) channels. The core contacts deliver the prime gamma-ray energy information and are digitized using four different gains to match the dynamic range required to meet the science goals. Linear Technology/Analog Devices LTC2194 ADCs provide the digitization for the science data. In addition, a fast low latency LTC2158-14 ADC on the core contact provides fast information for the trigger system. Firmware in the DM FPGA receives the ADC data and tags the samples with the timestamp. All downstream processing uses these time tags minimizing the need for latency tracking. A data stream of ~75Gbit/s raw ADC conversions flows on fiber optic links to a SFB for first level processing and data reduction. The data link utilizes 10-~10Gb/s fiber links and self-synchronizing 64b66b encoding to send the ADC data together with housekeeping information such as time stamp, temperatures and detector identification number. Individually floating 48V supplies generate the power for the DMs. The DMs generate all required regulated voltages locally. Fig. 3 shows a picture of the DM electronics.

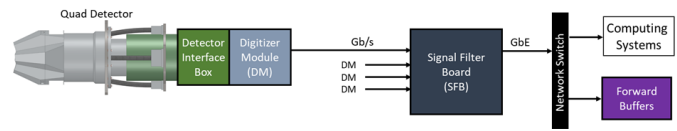

Fig. 1.  Schematic GRETA data acquisition hardware.

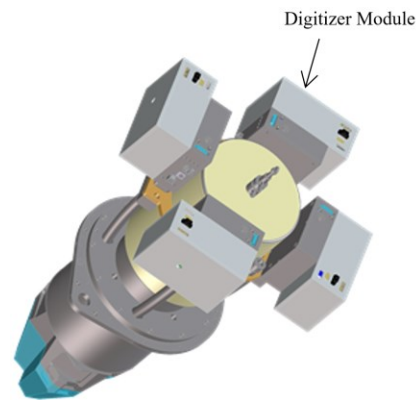

Fig. 2.  Detector module with 4 Digitizer Modules Mounted

One SFB receives a continuous stream of ADC data from four DMs to compute in real time the gamma-ray energy and interaction timestamp. Gamma-ray events, validated by the trigger system, are packaged into User Datagram Protocol (UDP) frames and sent over Ethernet to the Computing System Forward Buffers (FBs), reducing the data rate to an average of 32MByte/s/crystal. Computing Systems Forward Buffers, designed to handle the stateless UDP without dropping packets, receive the data from the SFB and buffer it to a computing cluster to pull for next level parameter extraction and data reduction.

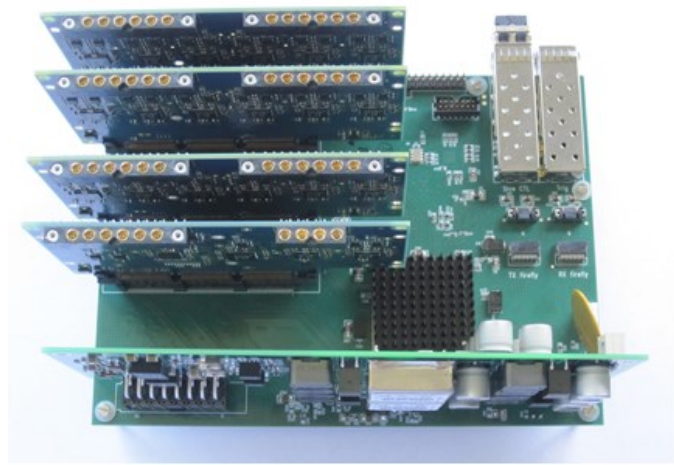

Fig. 3. Digitizer Module Electronics

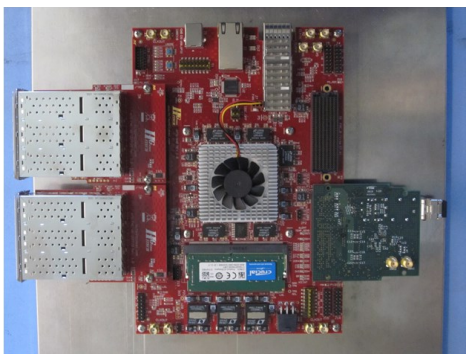

Fig. 4. Signal Filter Board

The SFB implements parallel real-time digital signal processing algorithms on 160 ADC channels from one Quad Detector Module.

The SFB implements parallel real-time digital signal processing algorithms on 160 ADC channels from a quad detector. A Xilinx Virtex Ultrascale FPGA provides the necessary high-speed receivers for the DM interfaces, the logic resources for the parallel digital signal processing, and the 25Gb/s Ethernet interface to computing. Commercial of the shelf hardware (Fig. 4) with plug in FMC modules comprises the SFB hardware. The main blocks of the digital signal processing consist of trapezoidal shaper to increase the signal to noise, high pass deconvolver (HPD) to revert the tail pulse generated by the charge sensitive preamplifiers bleeder resistor, baseline restoration to correct for numerical errors and nonlinearities, pulse height analyzer to extract the energy, pile-up rejection and leading edge discriminator. Table 1 shows the digital signal processing algorithms executed in the SFB.

## III. Performance

Performance of the DAQ hardware implementation and the SFB FPGA algorithms was verified using a GRETINA quad detector and controlled radiation sources. The firmware inside the SFB FPGAs executes the real-time signal processing on the signal waveforms path and provides small segments of the signals for readout. The availability of more powerful FPGAs motivated us to explore the details of the signal processing algorithms and functions to provide a deeper understanding of the operations and take advantage of capabilities in newer FPGAs. We built a set up that provided simultaneous transmission of DM data to a computer and an SFB. The offline data analysis performed by the computer cluster validated the newly designed DMs performance and provided validation of the energies values computed by the SFB. Fig. 5 demonstrates the DAQ energy resolution performance. The three curves show different trapezoidal filter shaping times, which are configurable in the SFB to optimize for the experimental gamma-ray rate conditions. Arrows show the GRETA requirements demonstrating the energy resolution we obtained is better than the requirement. At 1kCounts/s, the DAQ system achieved a FWHM of ~2.3keV, for a 10μs filter shaping time, compared to the required 2.5keV. At 50kCounts/s, the DAQ system achieved a FWHM of ~3.1KeV, for a 3μs filter shaping time, compared to the required 3.5keV.

Table 1. Digital Signal Processing inside the SFB

**Leading Edge Discrimination**
$y(n) = x(n) - x(n-k)$ (differentiation)
$y(n) = (x(n) + x(n-2)) + x(n-1) <<1$ (×4, Gaussian filtering)
Threshold comparison → LE discriminator time

**Constant Fraction Discrimination**
$y(n) = x(n) - x(n-k)$ (differentiation)
$y(n) = (x(n) + x(n-2)) + x(n-1) <<1$ (×2, Gaussian filtering)
$y(n) = x(n-k) <<ab - x(n)$ (constant fraction)
Zero crossing comparison → CFD time

**Trapezoidal filter and energy determination**
$y(n) = y(n-1) + ((x(n) + x(n-2m-k)) - (x(n-m) + x(n-m-k))$
Maximum tracking → energy

**Pole-Zero cancellation**
$l(n) = l(n-1) + x(n)$
$y(n) = x(n) + l(n)/t$ (where $t$ is the pre-amp time constant)

**Baseline Restoration**
$y(n) = BLR*y(n-1) + x(n)$ (×2)

## IV. Conclusion

In this paper, we have described the hardware and firmware implementation of the GRETA DAQ electronics. We have performed validation tests and they indicate that the design will meet the requirements. GRETA started the construction phase.

Tracking In-Beam Array (GRETINA)," IEEE Trans. Nuclear Science, Vol. 59, pp. 2494–2500, Oct. 2012.

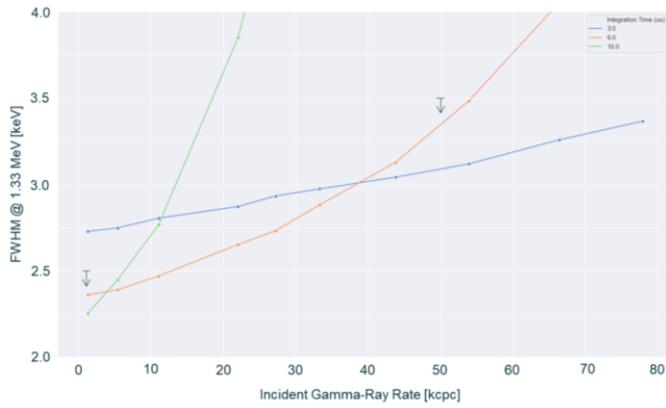

Fig. 5. Measure Energy Resolution using A GRETA Detector and a Digitizer Module Prototype. The three curves show the computed energy resolution for different filter shaping time settings (green: 10us, red: 6us, blue: 3us)